\documentclass[twocolumn,showpacs]{revtex4}
\usepackage{graphicx,epsf,bm}
\def\bR{\mbox{\boldmath $R$}}

\def\bk{\mbox{\boldmath $k$}}
\def\bp{\mbox{\boldmath $p$}}

\def\br{\mbox{\boldmath $r$}}
\def\bR{\mbox{\boldmath $R$}}

\begin{document}
\title{$^4$He energies and radii by the coupled-cluster method with
many-body average potential}
\author{M. Kohno\footnote{kohno@kyu-dent.ac.jp}}
\affiliation{Physics Division, Kyushu Dental College,
Kitakyushu 803-8580, Japan}
\author{R. Okamoto}
\affiliation{Senior Academy, Kyushu Institute of Technology, Kitakyushu 804-8550, Japan}

\begin{abstract}
The reformulated coupled-cluster method (CCM), in which average
many-body potentials are introduced, provides a useful
framework to organize numerous terms appearing in CCM equations,
which enables us to clarify the structure of the CCM theory and
physical importance of various terms more easily.
We explicitly apply this framework to $^4$He, retaining
one-body and two-body correlations as the first illustrating attempt.
Numerical results with using two modern nucleon-nucleon interactions
(AV18 and CD-Bonn) and their low-momentum interactions
are presented. The characters of short-range and many-body correlations
are discussed. Although not considered explicitly, the expression of
the ground-state energy in the presence of a three-nucleon force is given.
\end{abstract}

\maketitle

\section{Introduction}
The understanding of various properties of atomic nuclei starting from
the realistic nucleon-nucleon interactions is one of fundamental problems
in the theoretical nuclear physics. The recent studies in this field are
characterized by several new developments. Resulting from developments
through 1960's and 1990's, there are now various parameterizations
of the nucleon-nucleon interaction which reproduce experimental two-body
scattering and deuteron data with high precision. Systematic introduction
of a three-nucleon force is in progress especially in the framework of the
new development of the potential description in the chiral effective
field theory \cite{ME11,E06}. Progress is also seen in an effective
interaction theory in a restricted space, driven by active studies of
low-momentum interactions \cite{BOG}.
In addition, various ab-initio frameworks
of many-body calculations have been explicitly applied to atomic nuclei,
such as a Monte-Carlo method \cite{MC}, a no-core shell model \cite{NCSM}
and a coupled-cluster method (CCM) \cite{DH04}. Among them, the CCM is
a promising method toward heavier nuclei because of the advantage
of the size-extensivity.

The CCM was devised for many-body problems in nuclear physics
in 1950's \cite{C58,CK60} and the achievements in 1970's by the
Bochum group were reported in Ref. \cite{KLZ78}.
It was, however, almost discarded in nuclear physics community. The CCM
found its place in quantum chemistry \cite{BM07,SB09} as a tool of the
first-principle calculation. The method was reintroduced around 2000 in
the description of atomic nuclei \cite{HM99,DH04}. Because there are
specific difficulties in a description and a treatment of the short-range
part of a nucleon-nucleon interaction, applications of the CCM to nuclei
require deeper understanding of an effective interaction theory. It is not
surprising that a renewed interest arises in the CCM in parallel with the
development of low-momentum interaction theory \cite{BOG}.

There is already a considerable number of CCM calculations of light nuclei
in recent years \cite{DH04,KD04,HD07,HP09,HP10,JH11}.
These include the extension to excited states \cite{KD04,JH11},
nuclei far from the stability line \cite{HP09}
and heavier nuclei such as $^{40,48}$Ca \cite{HD07,HP10}.
Contributions of a three-nucleon force has also been estimated in this method \cite{HP07}.
Nevertheless, because of the usefulness of the CCM framework to solve the quantum
many-body problem as exactly as possible, it is important to clarify the structure
of this framework as much as possible and to try to obtain the physical understanding
of various correlations in a transparent way. Such an attempt was proposed by
Suzuki \cite{S92} in the early 1990's as an application of the similarity-transformation
theory for a quantum-mechanical eigenvalue problem. His reformulation
introduces average many-body potentials as the generalization of the one-body mean field.
The concept of the mean field is essentially important in almost all many-body
systems, as the independent single-particle model is empirically established
in those systems. As is shown below, the many-body average field is useful to
group various terms which appear in the CCM practical calculations, and thus is helpful
to elucidate the structure of the CCM many-body correlations. The transformation
of the Hamiltonian to the normal ordered form with respect to the reference state is
inherent in this formulation. Unfortunately, explicit
applications of this method to nuclei has not been carried out so far. It is useful at the
present stage to promote this formulation for the description of actual
nuclei with using the realistic nucleon-nucleon interaction, in view of the future
wide use of the CCM framework in nuclear many-body problems.

We recapitulate the CCM formulation with introducing average many-body potentials
in Sec. II, starting with the basic idea of the CCM. Although we do not consider the
contribution of a three-nucleon force in the present numerical calculations,
the expression including three-body terms are presented in the initial part
of Sec. II. As the first attempt of the actual application of this method to nuclei,
we employ the approximation in which only one-body and two-body amplitudes are
retained. This truncation is referred to as CCSD in the literature. Numerical
calculations are carried out for $^4$He with using two modern nucleon-nucleon potentials,
the CD-Bonn potential \cite{CDB} and the Argonne AV18 potential \cite{AV18}.
Results of the energy and radius of $^4$He in a harmonic oscillator model space
are presented in Sec. III. First, results with the bare potentials are shown,
and then the calculations with low-momentum equivalent interactions are discussed.
The magnitude of the contributions from the average many-body potential is estimated.
The independence of the results on the harmonic oscillator constant of the reference
state is demonstrated. If we do not include one-body amplitudes, the results
varies with changing the oscillator constant. Conclusions follow in Sec. IV.

\section{Coupled-cluster method}
We first give the idea of the CCM in its naive form, and next present the introduction
of many-body average potentials.
\subsection{Basic}
The basic ansats of the coupled-cluster method \cite{C58,CK60} is that the exact
state $|\Psi\rangle$ of the Hamiltonian $H$ of $A$ particles with the mass $m$ is
given by the transformation operator $e^S$ acting on the reference
state $|\Phi_0\rangle$:
\begin{equation}
 |\Psi\rangle = e^S|\Phi_0\rangle.
\end{equation}
The Hamiltonian $H$ consists of the one-body kinetic energy
$t_i=\frac{\hbar^2}{2m}\bk_i^2$, the two-nucleon interaction $v_{ij}$, the
three-nucleon interaction $v_{ijk}$, and so on. To be explicit, we write
the Hamiltonian as follows.
\begin{equation}
 H=\sum_i \frac{\hbar^2}{2m}\bk_i^2 - T_G +\sum_{i<j} v_{ij}+\sum_{i<j<k} v_{ijk},
\end{equation}
where  $T_G$ denotes the center-of-mass kinetic operator. In the following,
we absorb the one-body part of $T_G$ in the one-body kinetic energy,
namely the operator being $t_i=\frac{A-1}{A} \frac{\hbar^2}{2m} \bk^2$,
and the two-body part of $T_G$ in the two-body interaction $v_{ij}$.

Solving the time-independent Schr\"{o}dinger equation
\begin{equation}
He^S|\Phi_0\rangle =E_0 e^S|\Phi_0\rangle,\label{SHEQ}
\end{equation}
is to determine the transformation operator $S$, which consists of one-body part,
two-body part, and up to $A$-body part.
\begin{equation}
 S=\sum_i s_i+\sum_{i<j} s_{ij}+\sum_{i<j<k} s_{ijk}+ \cdots + s_{12\cdots A}.
\end{equation}
We refer to single-particle states in $|\Phi_0\rangle$ as hole states denoted by $h$
and other single-particle states as particle states denoted by $p$.
By definition, only the particle (hole) states appear in the bra (ket) configuration
of the transformation amplitude
$\langle p_1p_2\cdots p_n|s_{12\cdots n}|h_1h_2\cdots h_n\rangle_A$,
where the suffix $A$ means an antisymmetrized matrix element;
namely, $|h_1h_2\rangle_A\equiv |h_1h_2-h_2h_1\rangle$,
$|hh'h''\rangle_A\equiv |hh'h''-hh''h'+h'h''h-h'hh''+h''hh'-h''h'h\rangle$ and so on.

The prescription of the CCM is to rewrite the Schr\"{o}dinger equation (\ref{SHEQ}) as
\begin{equation}
 e^{-S}He^S|\Phi_0\rangle =E_0|\Phi_0\rangle.
\end{equation}
It is noted that in the standard CCM theory, the Hamiltonian is supposed to be
normal-ordered with respect to the reference state at this stage. Hence we denote
the Hamiltonian by $H_n$.
The operator $S$ is determined by the decoupling equation for
the similarity-transformed Hamiltonian $e^{-S}H_ne^S$:
\begin{equation}
 \langle np\:\mbox{-}\:nh |e^{-S}H_n e^S|\Phi_0\rangle=0,\label{eq:ccm}
\end{equation}
with $n=1,2,\cdots A$. Here, $\langle np\:\mbox{-}\:nh |$ represents an arbitrary
$n$-particle and $n$-hole state. Determining $S$, the energy is given by
\begin{equation}
 E_0=\langle \Phi_0|e^{-S}H_ne^S|\Phi_0\rangle.\label{eq:ccme}
\end{equation}
The similarity-transformation $e^{-S}H_ne^S$ may be evaluated systematically by the
Baker-Campbell-Hausdorff formula:
\begin{equation}
 e^{-S}H_ne^S = H_n+[H_n,S]+\frac{1}{2!}[[H_n,S],S]+\cdots .
\end{equation}
In practice, the Hamiltonian should be rearranged in the normal-order form
with respect to $|\Phi_0\rangle$ \cite{CZ66,DH04}. The transparent procedure of the
transformation in view of the effective interaction theory was formulated by Suzuki \cite{S92}.
This method with introducing many-body average potentials is outlined in the next subsection.

\subsection{Many-body average potential}
When the original Hamiltonian $H$ do not have more than three-nucleon
interactions, the expansion of $e^{-S}He^S$ terminates with the four-folded
commutator and produces various terms up to six-body operators.
If a three-nucleon interaction is present, more than five-folded commutators appear.
Suzuki and collaborators \cite{S92,SOK94} reformulated the CCM
by introducing many-body average fields, which enables us to write down the decoupling
equations in a compact form and thus to clarify the structure of various many-body
correlations.
We manipulate explicit expressions in this formulation and carry out numerical
calculations.

The similarity-transformed Hamiltonian may be reorganized by introducing an
auxiliary many-body potential
\begin{equation}
U=\sum_i u_i + \sum_{i<j} u_{ij}+\sum_{i<j<k}u_{ijk}+\cdots.
\end{equation}
The transformed Hamiltonian is written as
\begin{eqnarray}
  e^{-S}He^S&=& e^{-S}(H+U)e^S - e^{-S}Ue^S\nonumber \\
  &=& \{ \sum_i \tilde{h}_i+\sum_{i<j}\tilde{v}_{ij}+\sum_{i<j<k}\tilde{v}_{ijk}+\cdots\} \nonumber \\
  &-& \{ \sum_i \tilde{u}_i+\sum_{i<j}\tilde{u}_{ij}+\sum_{i<j<k}\tilde{u}_{ijk}+\cdots\}.
\end{eqnarray}
That is,
\begin{eqnarray}
 e^{-S}\{\sum_i t_i +\sum_{i<j} v_{ij}+\sum_{i<j<k} v_{ijk}\}e^S \nonumber \\
 = \sum_i (\tilde{h}_i-\tilde{u}_i)+\sum_{i<j}(\tilde{v}_{ij}-\tilde{u}_{ij}) \nonumber \\
 +\sum_{i<j<k}(\tilde{v}_{ijk}-\tilde{u}_{ijk})+\cdots .\label{eq:stf}
\end{eqnarray}
The prototype of the auxiliary potential for a many-body system is a Hartree-Fock
one. The one-body Hartree-Fock mean field for states $|a\rangle$ and $|b\rangle$
is defined by folding two-body interactions with respect to occupied states $|h\rangle$:
\begin{equation}
 \langle a|u_{HF}|b\rangle =\sum_h \langle ah|v_{12}|bh\rangle_A.
\end{equation}
The summation is depicted as a bubble insertion diagram of Fig. 1. The formulation
by Suzuki {\it et al.} \cite{S92,SOK94} generalize the HF potential to
many-body states. That is, the auxiliary potential $U$ is
determined so as to cancel all the bubble-insertion contributions.
\begin{figure}
\includegraphics[width=5cm]{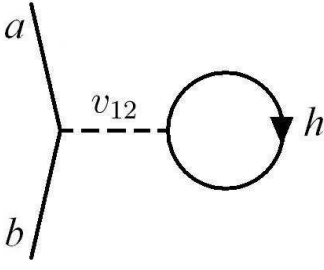}
\caption{Bubble insertion diagram for Hartree-Fock mean field.
The antisymmetrization is not shown explicitly.}
\end{figure}
\begin{widetext}
\begin{equation}
 \langle a_1\cdots a_n |\tilde{u}_{1\cdots n}|b_1\cdots b_n\rangle_A =
 \sum_{k\geq 1} \frac{(-1)^{k+1}}{k!} \sum_{h_1\cdots h_k}
 \langle a_1\cdots a_nh_1\cdots h_k|\tilde{v}_{1\cdots n+k}
 |b_1\cdots b_nh_1\cdots h_k\rangle_A,
\end{equation}
where $a$ and $b$ stand for either particle or hole state.
It is instructive to give explicit expressions for the one-body $\tilde{u}_1$ and
the two-body $\tilde{u}_{12}$.
\begin{eqnarray}
 \langle a |\tilde{u}_1|b\rangle &=& \sum_{h} \langle ah|\tilde{v}_{12}|bh\rangle_A
 -\frac{1}{2!}\sum_{hh'} \langle ahh'|\tilde{v}_{123}|bhh'\rangle_A
  +\frac{1}{3!}\sum_{hh'} \langle a_1a_2hh'|\tilde{v}_{1234}|b_1b_2hh'\rangle_A+\cdots, \\
  \langle a_1 a_2 |\tilde{u}_{12}|b_1b_2\rangle_A &=& \sum_{h}
 \langle a_1a_2h|\tilde{v}_{123}|b_1b_2h\rangle_A
 -\frac{1}{2!}\sum_{hh'} \langle a_1a_2hh'|\tilde{v}_{1234}|b_1b_2hh'\rangle_A
  +\cdots.
\end{eqnarray}
It is apparent that $\tilde{u}$ is a sum of bubble-insertion contributions of
the similarity-transformed interactions, if we rewrite the above definition of $\tilde{u}_1$ as
\begin{eqnarray}
 \langle a |\tilde{u}_1|b\rangle &=& \sum_{h} \langle ah|\tilde{v}_{12}-\tilde{u}_{12}
 |bh\rangle_A + \sum_{h} \langle ah|\tilde{u}_{12}|bh\rangle_A
  -\frac{1}{2!}\sum_{hh'} \langle ahh'|\tilde{v}_{123}|bhh'\rangle_A \\
 & & +\frac{1}{3!}\sum_{hh'h''} \langle ahh'h''|\tilde{v}_{1234}|bhh'h''\rangle_A+\cdots,
 \nonumber \\
 &=& \sum_{h} \langle ah|\tilde{v}_{12}-\tilde{u}_{12}|bh\rangle_A
  + \frac{1}{2!}\sum_{hh'} \langle ahh'|\tilde{v}_{123}|bhh'\rangle_A
  +\left(\frac{1}{3!}-\frac{1}{2!}\right)
 \sum_{hhh''} \langle ahh'h''|\tilde{v}_{1234}|bhh'h''\rangle_A+\cdots  \nonumber \\
  &=& \sum_{h} \langle ah|\tilde{v}_{12}-\tilde{u}_{12}|bh\rangle_A
  + \frac{1}{2!}\sum_{hh'} \langle ahh'|\tilde{v}_{123}-\tilde{u}_{123}|bhh'\rangle_A
 +\frac{1}{3!}\sum_{hhh''} \langle ahh'h''|\tilde{v}_{1234}|bhh'h''\rangle_A +\cdots,
\end{eqnarray}
and also for $\tilde{u}_{12}$
\begin{eqnarray}
  \langle a_1 a_2 |\tilde{u}_{12}|b_1b_2\rangle_A &=& \sum_{h} \{
 \langle a_1a_2h|\tilde{v}_{123}-\tilde{u}_{123}|b_1b_2h\rangle_A
 +\langle a_1a_2h|\tilde{u}_{123}|b_1b_2h\rangle_A\}
 -\frac{1}{2!}\sum_{hh'} \langle a_1a_2hh'|\tilde{v}_{1234}|b_1b_2hh'\rangle_A
  +\cdots \nonumber \\
 &=& \sum_{h} \langle a_1a_2h|\tilde{v}_{123}-\tilde{u}_{123}|b_1b_2h\rangle_A
 +\frac{1}{2!}\sum_{hh'} \langle a_1a_2hh'|\tilde{v}_{1234}|b_1b_2hh'\rangle_A
  +\cdots \label{eq:u12}.
\end{eqnarray}
\end{widetext}
Therefore, $e^{-S}(H+U)e^S$ is now arranged
in the normal-ordered form with respect to the reference state. It means that the
following decoupling conditions should be imposed.
\begin{equation}
 \langle np\:\mbox{-}\:nh|e^{-S}(H+U)e^S|\Phi_0\rangle=0.
\end{equation}
This corresponds to considering the decoupling condition for the
normal-ordered Hamiltonian $H_n$ in the standard CCM, Eq. (\ref{eq:ccm}).
The potential $U$ thus defined may be called a many-body average potential.

To proceed to actual calculations, explicit expressions of the similarity-transformed
interaction $e^{-S}He^S$ are needed. The result of the straightforward calculation
is given in the Appendix. In these expressions the terms including a three-nucleon
interaction and a three-body amplitude $s_{ijk}$ are retained. However, because
we do not consider these contributions in the CCSD truncation in this paper,
we do not keep these terms here after.

\subsection{Decoupling equations and energy}
Some manipulation is necessary to obtain explicit expressions of matrix
elements of $\tilde{u}$ in terms of the original interactions $v_{12}$ and the
correlation amplitudes $s_{1}$ and $s_{12}$. Because
$\tilde{h}_1=e^{-s_1}t_1e^{s_1}+\tilde{u}_1= (1-s_1)t_1(1+s_1)+\tilde{u}_1$,
the one-particle-one-hole decoupling equation becomes
\begin{equation}
 \langle p|\tilde{h}_1|h\rangle =\langle p|(1-s_1)t_1(1+s_1)+\tilde{u}_1|h\rangle=0.
\end{equation}
The explicit expression of $\langle p|\tilde{u}_1|h\rangle$ is given below,
Eq. (\ref{eq:phu1}).
Next, the two-particle-two-hole decoupling equation
$\langle p_1p_2|\tilde{v}_{12}|h_1h_2\rangle_A=0$ reads
\begin{eqnarray}
 & & \langle p_1p_2|\tilde{v}_{12}-\tilde{u}_{12}+\tilde{u}_{12}|h_1h_2\rangle_A \nonumber\\
 & & =\langle p_1p_2|(1-s_1-s_2+s_1s_2-s_{12})v_{12} \nonumber\\
  & & \times (1+s_1+s_2+s_1s_2+s_{12})
 +\tilde{u}_{12}|h_1h_2\rangle_A=0.\hspace*{5mm} \label{eq:refccm}
\end{eqnarray}
Figuring out the explicit form of the matrix
element $\langle p_1p_2|\tilde{u}_{12}|h_1h_2\rangle_A$ according to Eq. (\ref{eq:u12}),
this equation can be written in the following rather compact manner:
\begin{widetext}
\begin{eqnarray}
 & & \langle p_1p_2|(1-s_1-s_2+s_1s_2+s_{12})v_{12}
 (1+s_1+s_2+s_1s_2+s_{12})|h_1h_2 \rangle_A \nonumber\\
 & & + \langle p_1p_2|(\epsilon_1 +\epsilon_2)s_{12}
-s_{12}(\epsilon_1+\epsilon_2)|h_1h_2\rangle_A
   + \langle p_1p_2|u_{12}^A+u_{12}^B| h_1h_2 \rangle_A=0\label{eq:dc12},
\end{eqnarray}
where
\begin{eqnarray}
 \langle h_1|\epsilon_1|h_2\rangle &\equiv& \langle h_1|t_1(1+s_1)+\tilde{u}_1|h_2\rangle, \\
 \langle p_1|\epsilon_1|p_2\rangle &\equiv& \langle p_1| (1-s_1)t_1
 +\tilde{u}_1|p_2\rangle, \\
 \langle h_1|\tilde{u}_1|h_2\rangle &=&\sum_{h'}\langle h_1h'|
 v_{12}(1+s_1+s_2+s_1s_2 +s_{12})|h_2h'\rangle_A \label{eq:u1h}, \\
  \langle p_1|\tilde{u}_1|p_2\rangle &=& \sum_{h'}
 \langle p_1h'|(1-s_1)v_{12}(1+s_1+s_2) |p_2h'\rangle_A 
 -\sum_{p'} \langle p_1p'|s_{12}v_{12}|p_2p'\rangle_A \label{eq:u1p},\\
 \langle p| \tilde{u}_1|h\rangle & =& \langle ph' |t_2s_{12}
 +(1-s_1)v_{12}(1+s_1+s_2+s_1s_2+s_{12})| hh' \rangle_A \nonumber\\
 & & + \frac{1}{2}\sum_{h'h''}\sum_{p'}\{2\langle h'h''|v_{12}|p'h''\rangle_A
 \langle pp'|s_{12}|hh'\rangle_A+\langle h'h''|v_{12}|p'h\rangle_A
 \langle pp'|s_{12}|h'h''\rangle_A\} \nonumber\\
 & & +\frac{1}{2}\sum_{h'h''} \frac{1}{2}\sum_{p'p''}2\langle h'h''|v_{12}|p'p''\rangle_A
 \{2\langle pp'|s_{12}|hh'\rangle_A \langle p''|s_1|h''\rangle
   + \langle pp'|s_{12}|h'h''\rangle_A \langle p''|s_1|h\rangle \}\label{eq:phu1},\\
\langle p_1p_2|u_{12}^A|h_1h_2\rangle_A &\equiv& \sum_{h'p'}
 \left[ \{\langle p_2h'|(1-s_1)v_{12}|p'h_1\rangle_A
 +\sum_{p''}\langle p_2h'|(1-s_1)v_{12}|p'p''\rangle_A \langle p''|s_1|h_1\rangle\}
  \langle p_1p'|s_{12}|h_2h'\rangle_A \right. \nonumber\\
 & & -\{\langle p_2h'|(1-s_1)v_{12}|p'h_2\rangle_A
 +\sum_{p''}\langle p_2h'|(1-s_1)v_{12}|p'p''\rangle_A \langle p''|s_1|h_2\rangle\}
  \langle p_1p'|s_{12}|h_1h'\rangle_A\nonumber\\
 & & +\{\langle p_1h'|(1-s_1)v_{12}|p'h_1\rangle_A
 +\sum_{p''}\langle p_1h'|(1-s_1)v_{12}|p'p''\rangle_A \langle p''|s_1|h_1\rangle\}
  \langle p'p_2|s_{12}|h_2h'\rangle_A\nonumber\\
 & & \left. -\{\langle p_1h'|(1-s_1)v_{12}|p'h_2\rangle_A
 +\sum_{p''}\langle p_1h'|(1-s_1)v_{12}|p'p''\rangle_A \langle p''|s_1|h_2\rangle\}
  \langle p'p_2|s_{12}|h_1h'\rangle_A \right],  \label{eq:u12A}\\
\langle p_1p_2|u_{12}^B|h_1h_2\rangle_A &\equiv& 
 4\frac{1}{2}\sum_{h'h''}\frac{1}{2}\sum_{p'p''} \langle h'h''|v_{12}|p'p''\rangle_A
 \{\langle p_1p'|s_{12}|h_1h'\rangle_A \langle p_2p''|s_{12}|h_2h''\rangle_A\nonumber\\
 & & \hspace*{50mm}-\langle p_1p'|s_{12}|h_2h'\rangle_A
 \langle p_2p''|s_{12}|h_1h''\rangle_A\} \label{eq:u12B}.
\end{eqnarray}
\end{widetext}

\noindent
Note that $\frac{1}{2}\sum_{hh'} |hh'\rangle_A \langle hh'|$
($\frac{1}{2}\sum_{pp'} |pp'\rangle_A \langle pp'|$) is a unit operator
in an occupied (unoccupied) space. The $s_{12}$ contribution to the
single-particle potential $\tilde{u}_1$, Eqs. (\ref{eq:u1h}) and (\ref{eq:u1p}),
is diagrammatically represented as in Fig. 2, and the potentials $u_{12}^A$ and $u_{12}^B$,
Eqs. (\ref{eq:u12A}) and (\ref{eq:u12B}), in Fig. 3. It can be checked that Eq. (\ref{eq:refccm})
completely agrees with the CCSD decoupling equation in the
literature, as it should: e.g., Eq. (35) of ref. \cite{DH04}.

\begin{figure}
\includegraphics[width=8cm]{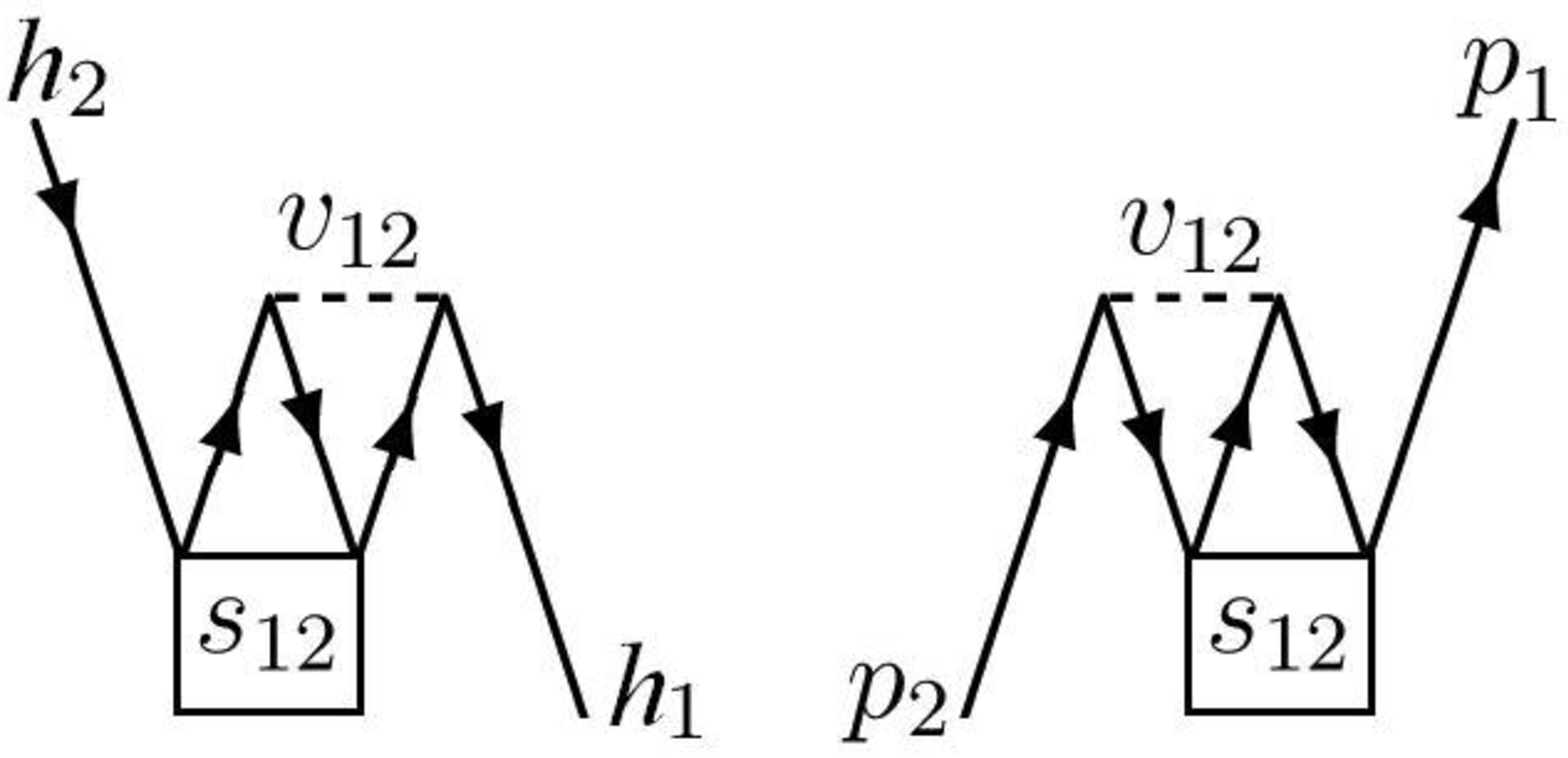}
\caption{Diagram representation of the $s_{12}$ contribution to the
one-body average potential $\tilde{u}_1$, Eqs. (\ref{eq:u1h}) and (\ref{eq:u1p})
The antisymmetrization is not shown explicitly.}
\bigskip
\includegraphics[width=8cm]{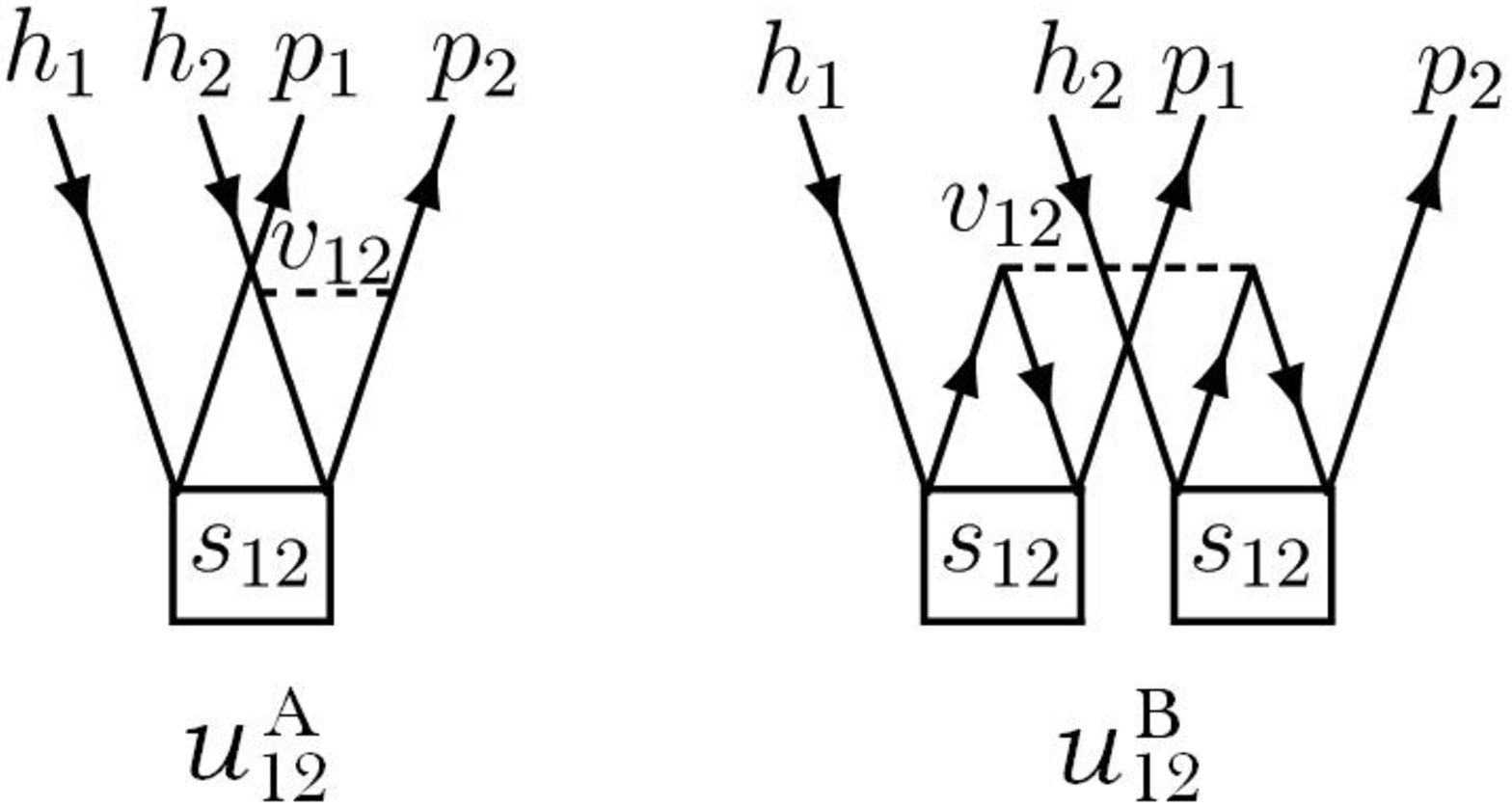}
\caption{Diagram representation of the $s_{12}$ contribution to the two-body
average potentials $u_{12}^A$ and $u_{12}^B$, Eqs. (\ref{eq:u12A}) and (\ref{eq:u12B}),
respectively. The antisymmetrization is not shown explicitly.}
\end{figure}

We need more explicit expressions to evaluate matrix elements such
as $\langle p_1p_2|(s_1+s_2)v_{12}|h_1h_2\rangle_A$, and algebraic calculations
of the angular-momentum coupling for the numerical calculations
of these expressions. We solve these non-linear equations for $s_i$ and $s_{ij}$
in an iterative way, using essentially the Newton method.

\subsection{Energy and root-mean-square radius}
The ground-state energy is obtained by Eq. (\ref{eq:ccme}). In the case that
only two-body interactions are present, amplitudes higher than three-body
$s_{ijk}$ do not contribute to the CCM energy and the expression becomes:
\begin{eqnarray}
 E_0&=&\sum_{h} \langle h|t_1(1+s_1)|h\rangle
 +\frac{1}{2}\sum_{hh'} \langle hh'|v_{12}\nonumber \\
 & & \times (1+s_1+s_2+s_1s_2+s_{12})|hh'\rangle_A.
\end{eqnarray}
Effects of the higher order correlations enter through
the one-body and two-body decoupling conditions. Although we do
not consider contributions of a three-nucleon force in this article,
it is instructive to present the energy expression in the presence
of $v_{ijk}$ and $s_{ijk}$, which is given in Appendix D. There we
further rewrite the expression by introducing an effective two-nucleon interaction
obtained from a three-nucleon interaction. 

In the CCM framework, the evaluation of expectation
values $\langle\Psi|Q|\Psi\rangle$ of  other
observables $Q$ than the energy needs explicit expansion of $e^{S^\dagger}Qe^S$.
Although only linked-cluster diagrams are needed to be taken into account,
a perturbative estimation in all-orders is not feasible. If the energy is determined
variationally, we may use the Feynmann-Hellman theorem to obtain the
matrix element of the given observable $Q$ by solving the problem
of the constrained Hamiltonian $H+\lambda Q$. That is, supposing the state
$|\Psi(\lambda)\rangle$ to be the solution of the forced Hamiltonian, the
following relation holds.
\begin{equation}
 \langle \Psi|Q|\Psi\rangle = \lim_{\lambda\rightarrow 0} \frac{d}{d\lambda}
 \langle \Psi (\lambda)|H+\lambda Q|\Psi(\lambda)\rangle. \label{eq:exv}
\end{equation}
Because the CCSD energy does not have a rigorous variational property,
 $\langle \Psi(\lambda)|H|\Psi(\lambda)\rangle$ may depend linearly
on $\lambda$, which introduces some uncertainty to the calculated expectation
value of $Q$. Nevertheless, we calculate the root-mean-square (r.m.s.) radius of
the $^4$He by this method, which is a square root of the expectation
value of the operator
\begin{eqnarray}
 Q_r &=& \frac{1}{A} \sum_{i=1}^A (\br_i-\bR)^2 \nonumber \\
  &=& \frac{1}{A}\left(1-\frac{1}{A}\right)\sum_{i=1}^A r_i^2
 - \frac{2}{A^2}\sum_{i<j} \br_i\cdot \br_j ,
\end{eqnarray}
where $\bR\equiv \frac{1}{A}\sum_i \br_i$ is the center-of-mass coordinate.
Because the CCSD energy reaches almost variational limit in practice,
the estimation by the constrained calculation is more reliable than
the perturbative estimation in low-orders. As is shown explicitly in the next section,
the reliability of the prescription of Eq. (\ref{eq:exv}) is suggested by the fact that
the calculated r.m.s. radii are almost independent on the oscillator constant
of the harmonic oscillator basis. 

It is helpful to note that the r.m.s. radius of the $^4$He reference state
$|\Phi_0\rangle\equiv |(0s)^4\rangle$ with the oscillator constant $\nu$
is given by
\begin{equation}
 \sqrt{\langle \Phi_0|Q_r|\Phi_0\rangle}=\frac{3}{2\sqrt{2\nu}}. \label{eq:hor}
\end{equation}

\section{Results}
We carry out numerical calculations in the harmonic oscillator
basis. The oscillator constant is denoted by $\nu$ which is related
to the frequency $\omega$ as $\hbar \omega =\frac{\hbar^2}{m}\nu$
with $m$ being the nucleon mass. The size of the harmonic-oscillator
basis is specified by the quantum number $N_{max}=(2n+\ell)_{max}$,
where $n$ is a nodal quantum number ($n=0,1,2\cdots$) and $\ell$
an orbital angular momentum. The number of the major shells included
is given by $N_{max}+1$. To see the dependence of CCM calculations
on the size of the model space, we use four different choices
of $N_{max}$, as shown in Fig. 4.

As the first application of the reformulated CCM, we consider only $^4$He,
for which the reference state $|\Phi\rangle$ is taken to be $(0s)^4$.
To keep the computaional cost low, we set $N_{max} \leq 22$
and introduce the cutoff at the upper right corner of the model space as in Fig. 4.
Two-body partial waves up to $J=6$ are included. We employ two realistic
nucleon-nucleon interactions; the CD-Bonn potential \cite{CDB}
and the AV18 potential \cite{AV18}. The charge dependence of
the CD-Bonn potential is averaged in actual calculations for the use
of the isospin basis. These two potentials are different
in their strength of the tensor force. The former is known to have
relatively weak tensor components. Discussions are chiefly presented
for the results of the CD-Bonn potential.

\begin{figure}
\includegraphics[width=6cm]{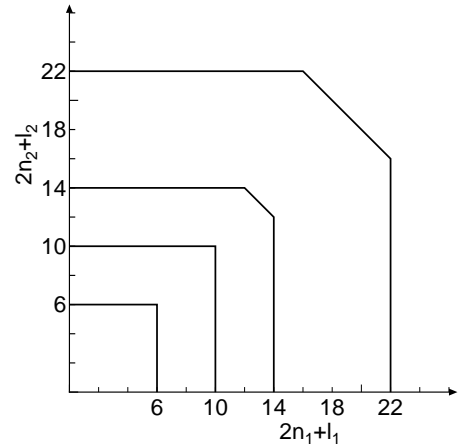}
\caption{Harmonic-oscillator model space for CCM. Four choices,
$(2n+\ell)_{max}=6, 10, 14$, and 22, are used for numerical calculations.}
\end{figure}

Two-body matrix elements of the bare nucleon-nucleon interaction
in the harmonic oscillator basis are evaluated by the Talmi-Moshinsky
transformation. Because the computation of Talmi-Moshinsky coefficients
is time consuming when the oscillator quantum number becomes large,
we use the approximation that the center-of mass (CM) quantum number
is restricted to be $(2n+\ell)_{CM}\le 6$. We have checked that the error
due to this simplification is negligibly small.
 
\subsection{Calculations with CD-Bonn bare interaction}
Calculated CCSD ground-state energies of $^4$He with the CD-Bonn potential
are shown in Fig. 5, as a function of the size of the harmonic-oscillator base,
$N\equiv 2n+\ell$, with the oscillator constant $\nu=0.56$ fm$^{-2}$.
As a reference, the energy of the Fadeev-Yakbovsky calculation \cite{NOG02}
with the same interaction is also shown. Increasing the size of the model
space, the CCSD energy goes down toward to the F-Y energy. However,
at $N_{max}=22$, the energy is still short by about
8 MeV to the reference value of the Faddeev-Yakubovsky calculation.
When the AV18 potential is employed, the shortage is about 21 MeV,
compared with the corresponding F-Y result of -24.3 MeV.
This result is not surprising, because the s.p. space is not sufficient to
properly treat the short-range correlation of the ordinary realistic
nucleon-nucleon interaction.
The momentum scale in the model space may be estimated 
by the expectation value of the square of the momentum operator
$\langle \bp^2\rangle=(N_{max}+\frac{3}{2})\nu$. The number
is $\sqrt{\langle \bp^2\rangle}=3.6$ fm$^{-1}$ for $N_{max}=22$
and $\nu=0.56$ fm$^{-2}$, which is much smaller than the high momentum
relevant to the nucleon-nucleon short-range correlation.

\begin{figure}
\includegraphics[width=7cm]{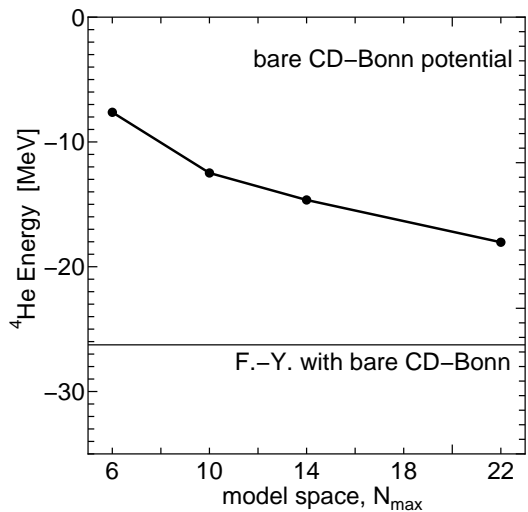}
\caption{Dependence of CCSD ground-state energies of $^4$He
on the harmonic-oscillator model-space size $N_{max}$ with the bare
CD-Bonn potential \cite{CDB}. The oscillator constant
is $\nu=0.56$ fm$^{-2}$ ($\hbar\omega=23.2$ MeV).
The energy by the Faddeev-Yakubovsky calculation \cite{NOG02}
is shown for reference. 
}
\end{figure}

It is difficult in practice to employ a large harmonic oscillator basis
for CCM calculations enough to deal with the nucleon-nucleon short-range
correlation. The established wisdom in nuclear physics is to treat the short range
singularity by solving the $G$-matrix equation as a two-body problem
in the nuclear medium \cite{BLM54,BR67}. Though the perturbative expansion in terms
of $G$-matrices is a solid framework, the CCM provides a more systematic
and compact way to treat higher-order correlations. One standard approximation
besides the $G$-matrix approach is to use a two-step method, in which
two-body correlations are initially treated in a large space and then solve
many-body correlations in a restricted space. This strategy is adopted,
e.g., in the unitary-model-operator approach \cite{SO94}.

A different method has been recently developed. Namely, the
bare interaction is transformed to the half-on-shell two-body
equivalent interaction in low-momentum space \cite{BOG}, which is
free from the high-momentum singularity. This interaction has been
shown to be used in perturbation calculations in a low-momentum space.
Though the linkage between the harmonic oscillator basis and
the low-momentum space is somewhat obscure and therefore the
treatment in the harmonic oscillator basis from the beginning is desirable,
we use the low-momentum space interaction in this paper
because of its simplicity and possibilities of the comparison with
calculations in the literature \cite{HD07}.

It is noted that in a restricted model space more than three-body
correlations involving high-momentum states are totally dropped.
These contributions are to be recovered by more than
three-nucleon induced interactions in a restricted space. Otherwise
some adjustable parameters may be introduced, if we take a viewpoint
that high-momentum components of bare nucleon-nucleon
interactions are themselves cannot be determined without uncertainties.

\subsection{Calculations with low-momentum interaction}
In this subsection, calculated results for low-momentum space
equivalent interactions of three cutoff momenta, $\Lambda=4.0$, $3.0$,
and $1.9$ fm$^{-1}$, are presented. The CCSD $^4$He ground-state
energies and radii with the CD-Bonn potential are shown in Figs. 6
and 7 as a function of the size of the harmonic-oscillator basis.

\begin{figure}
\includegraphics[width=7cm]{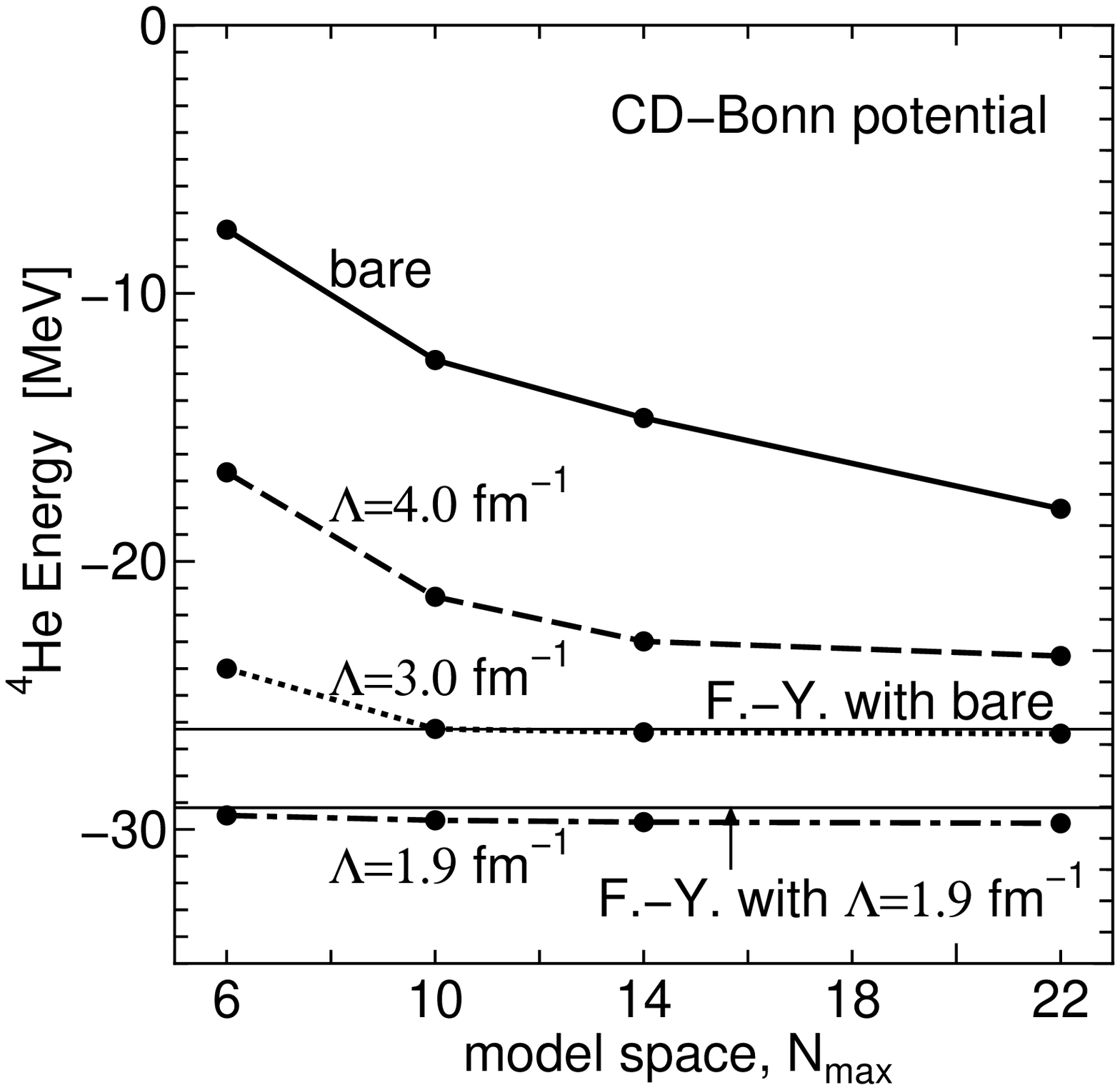}
\caption{Dependence of CCSD ground-state energies of $^4$He
on the harmonic-oscillator model-space size $(2n+\ell)_{max}$
with low-momentum equivalent interactions of three cutoff momenta,
$\Lambda=4.0$, $3.0$, and $1.9$ fm$^{-1}$. The bare interaction is
the CD-Bonn potential and the oscillator constant
is $\nu=0.56$ fm$^{-2}$ ($\hbar\omega=23.2$ MeV).
}
\medskip
\includegraphics[width=7cm]{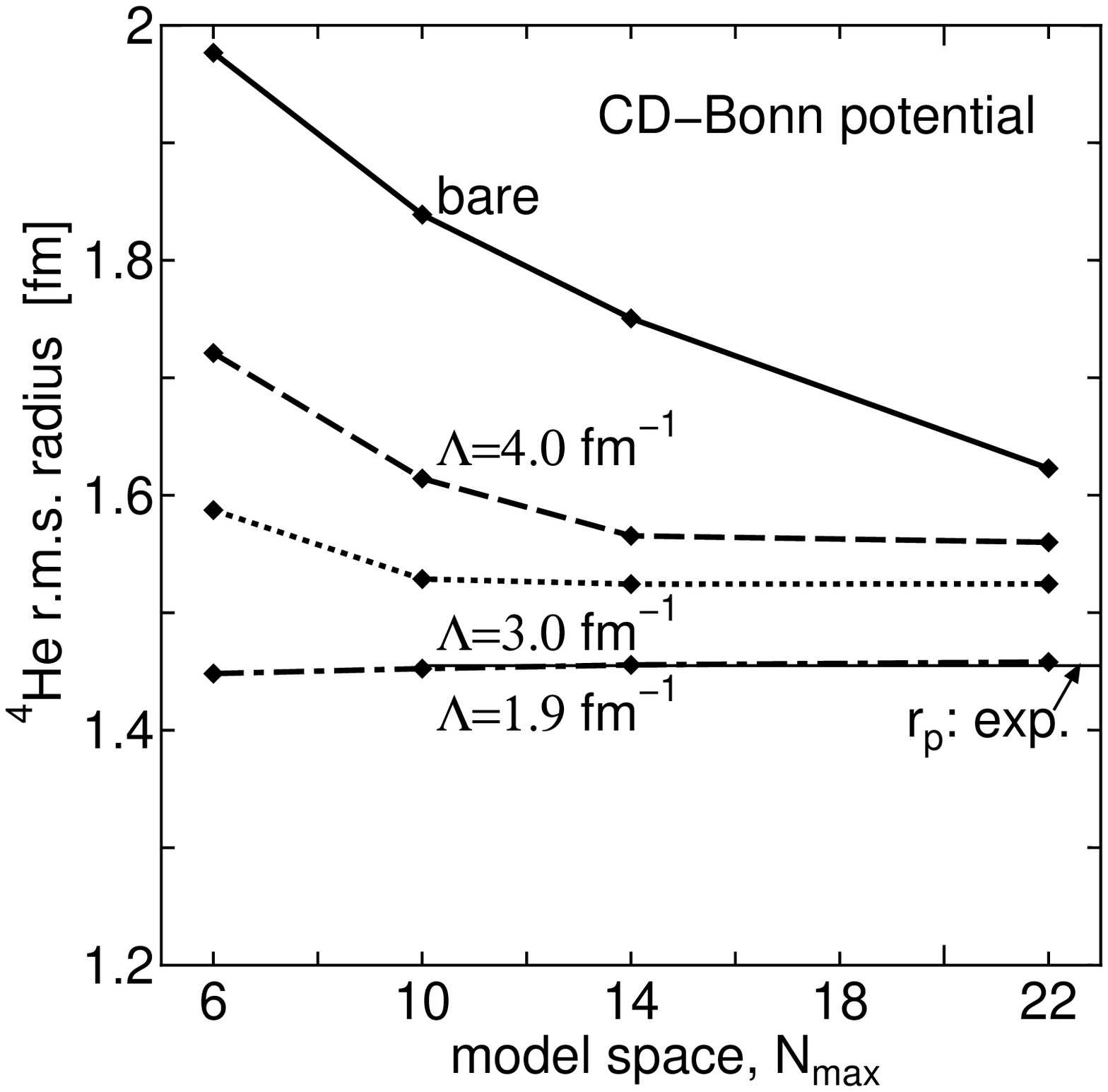}
\caption{Same as in Fig. 6, but for CCSD ground-state point-nucleon r.m.s.
radii of $^4$He. The experimental point proton radius is shown as a guide,
which is taken from ref. \cite{S08}.}
\end{figure}

Results with $\Lambda=1.9$ fm$^{-1}$ show that the energy of
the Faddeef-Yakubovsky calculation with the same low-momentum
equivalent interaction is almost reproduced already at $N_{max}=6$.
This implies that if the two-body correlations are renormalized
at $\Lambda=1.9$ fm$^{-1}$, more than three-body correlations in
the model space scarcely change energy. The evaluated radius is
also insensitive to the size of the model space
for $\Lambda =1.9$ fm$^{-1}$. On the other hand, the difference
between the energy with $\Lambda=1.9$ fm$^{-1}$ and the F-Y energy
with the bare interaction suggests that more than three-body
correlations involving higher-momentum components are of the order
of a few MeV. Observing that the energy at $\Lambda=3$ fm$^{-1}$
coincides with the F-Y energy, we see that the three-body
correlations at the momentum range of $2\sim 3$ MeV/c is
not negligible. It is remarked that the energy with
$\Lambda=1.9$ fm$^{-1}$ is slightly lower than the F-Y energy
with the corresponding low-momentum interaction. This feature
is different from the tendency in ref. \cite{HD07}.

\begin{figure}
\includegraphics[width=7cm]{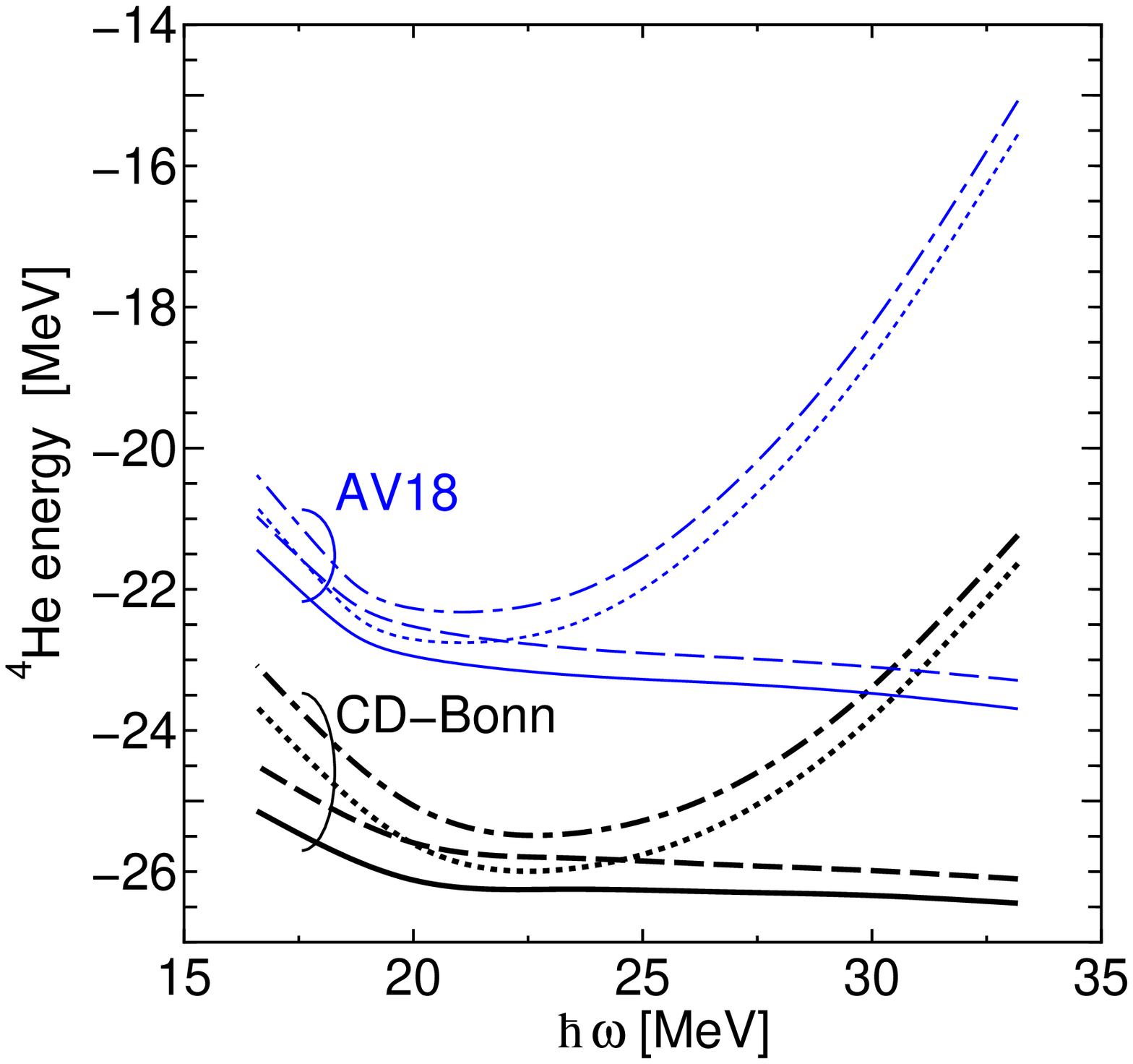}
\caption{Oscillator constant dependence of the ground-state
energies of $^4$He for the model space $N_{max}=10$ with using the
low-momentum equivalent interaction of $\Lambda=3$ fm$^{-1}$.
The solid curve is the result of full CCSD calculation. The dashed curve
shows the result with discarding $u_{12}^A$ and $u_{12}^B$. The dotted
and dot-dashed curves are results of the calculation for $s_1=0$
with and without $u_{12}^A$ and $u_{12}^B$, respectively.
}
\bigskip

\includegraphics[width=7cm]{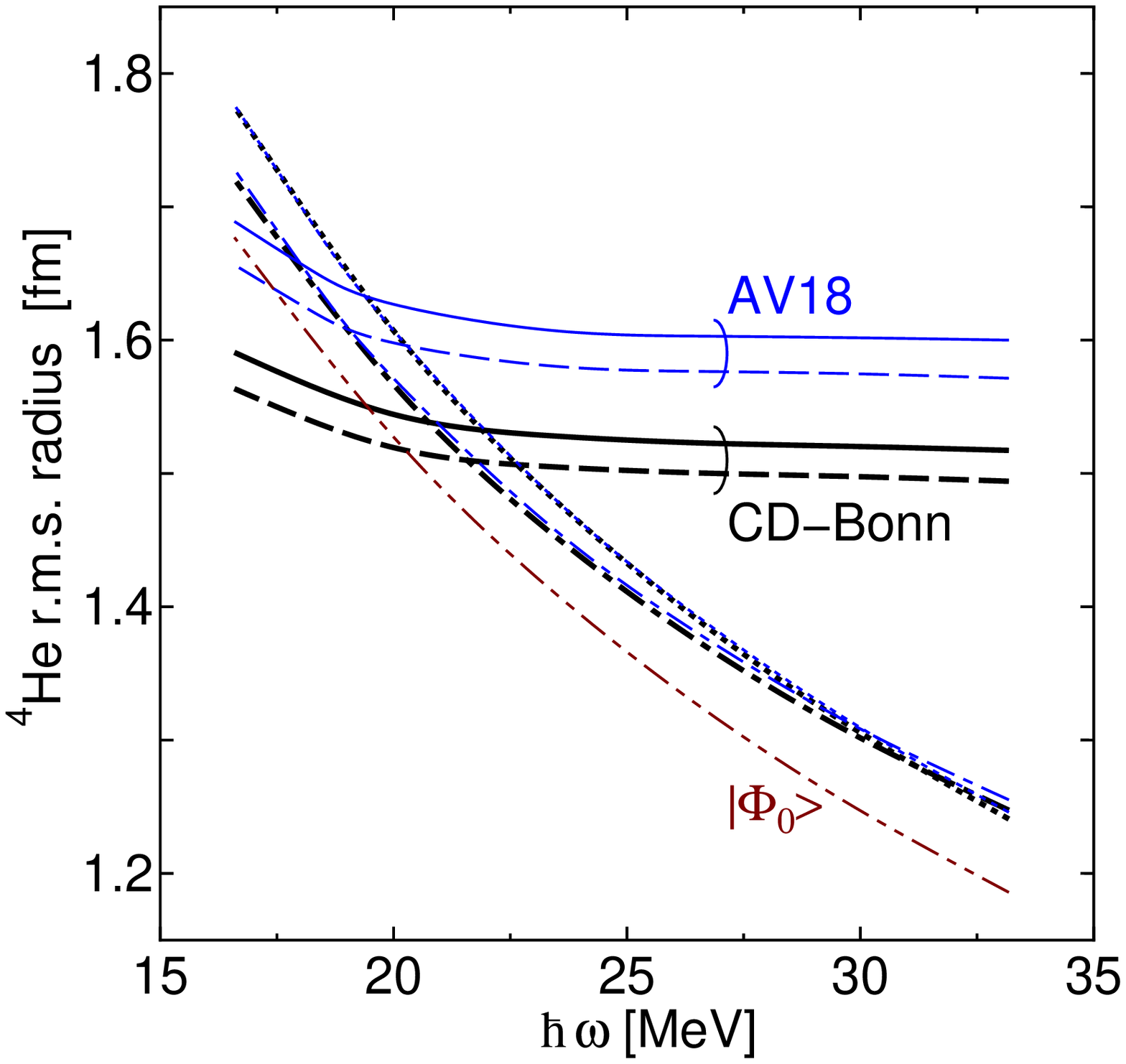}
\caption{Same as in Fig. 8, but for ground-state point-nucleon r.m.s.
radii of $^4$He. The thin dot-dashed curve is almost indistinguishable from
the bold dot-dashed curve. The r.m.s. radius of the reference state
$|\Phi_0\rangle$, Eq. (\ref{eq:hor}), is shown by a two-dot-dashed curve.}
\end{figure}

\begin{figure}
\includegraphics[width=7cm]{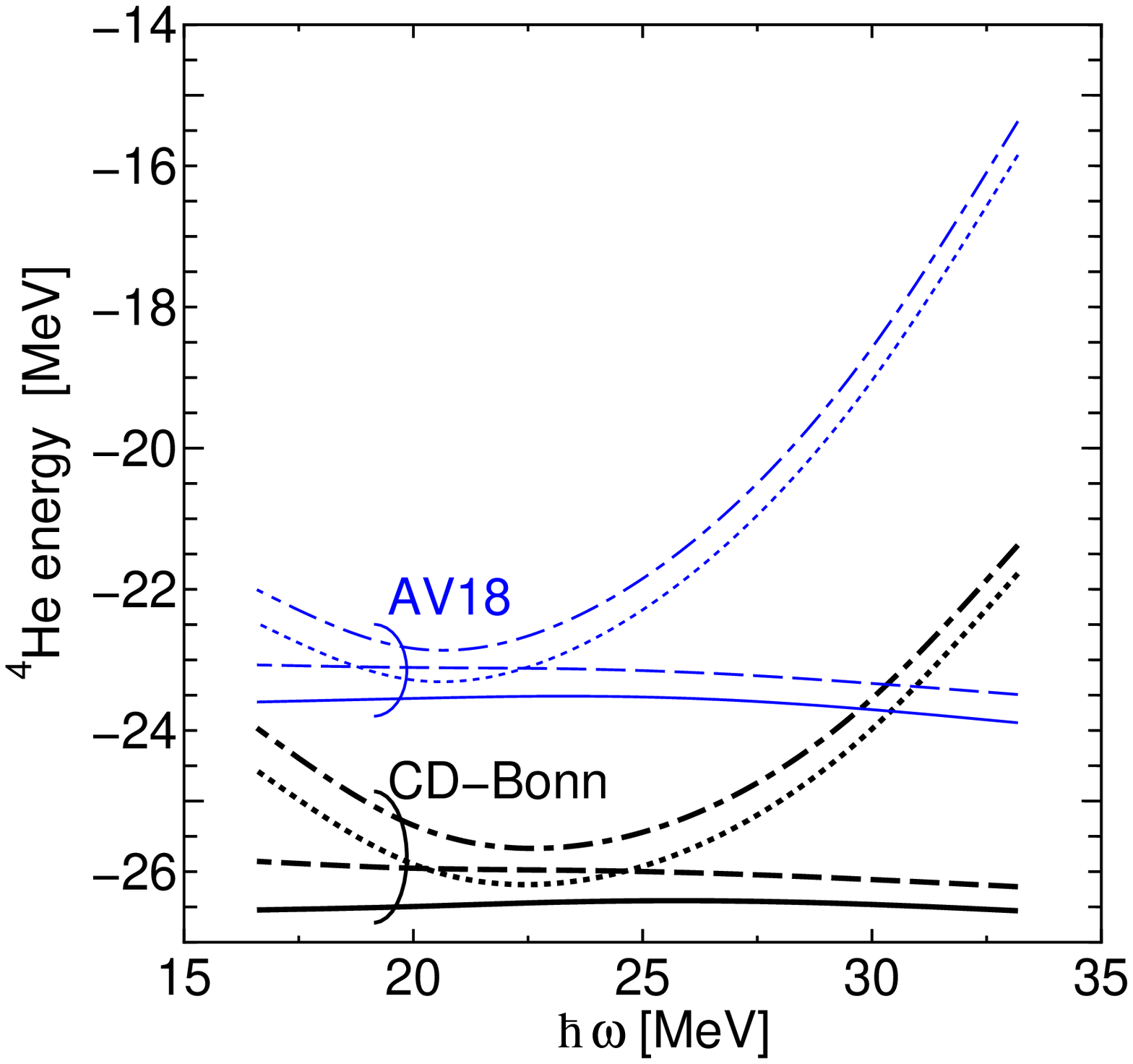}
\caption{Oscillator constant dependence of the ground-state energies
of $^4$He for the model space $N_{max}=22$ with using the low-momentum
equivalent interaction of $\Lambda=3$ fm$^{-1}$.
The solid curve is the result of full CCSD calculation. The dashed curve
shows the result with discarding $u_{12}^A$ and $u_{12}^B$. The dotted and
dot-dashed curves are results of the calculation for $s_1=0$ with and
without $u_{12}^A$ and $u_{12}^B$, respectively.
}

\bigskip
\includegraphics[width=7cm]{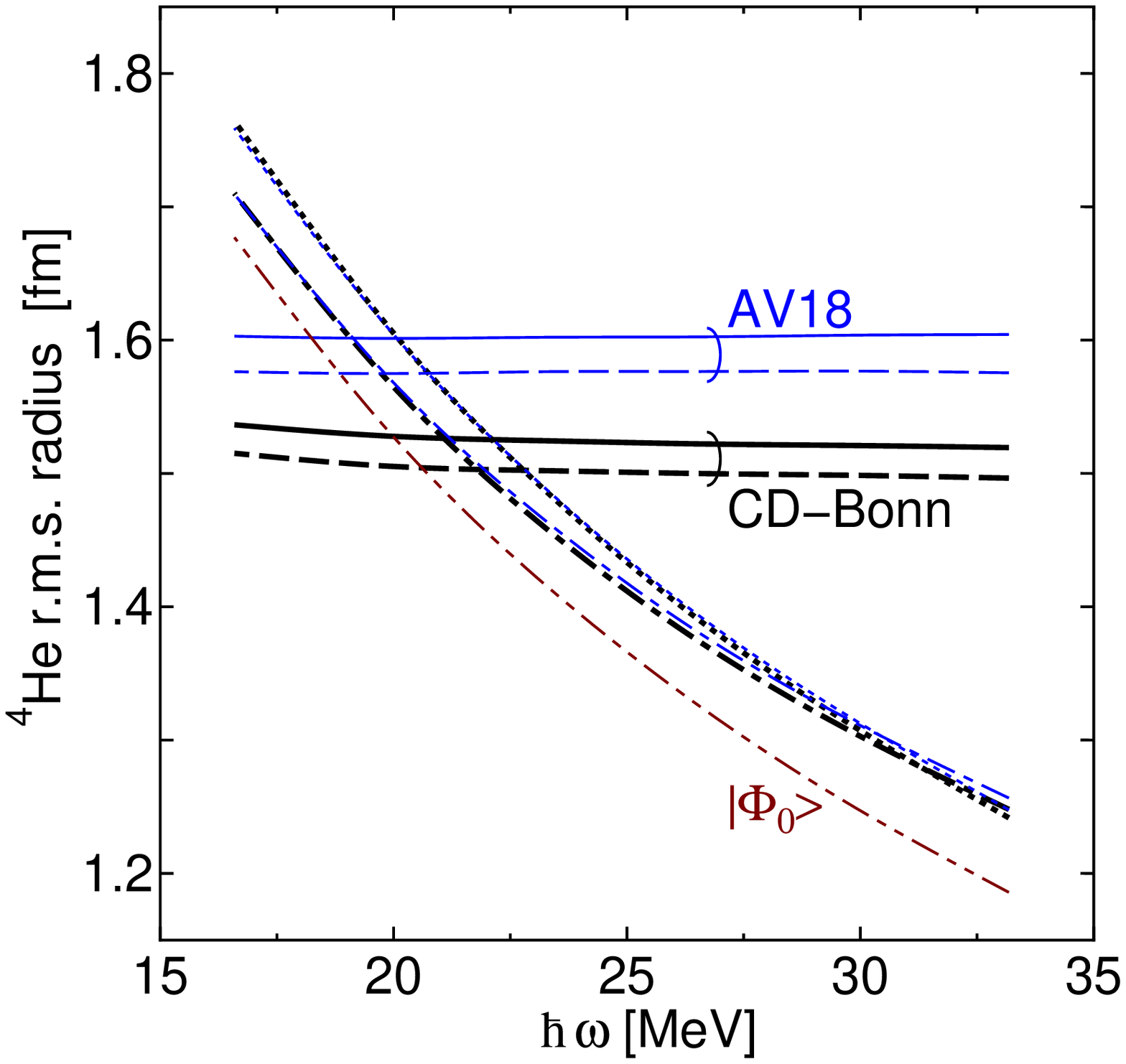}
\caption{Same as in Fig. 10, but for ground-state point-nucleon r.m.s.
radii of $^4$He. The thin dot-dashed curve is almost indistinguishable from
the bold dot-dashed curve. The r.m.s. radius of the reference state
$|\Phi_0\rangle$, Eq. (\ref{eq:hor}), is shown by a two-dot-dashed curve.}
\end{figure}

Next, we turn to the oscillator constant dependence of the calculated
results for CD-Bonn and AV18. This case, we also show results of the
calculation in which $s_1$ is set to be 0 and/or the potentials $u_{12}^{A,B}$
in Eq. (\ref{eq:refccm}) are omitted. Figs. 8 and 9 are the results for the model space
$N_{max}=10$ , and Figs. 10 and 11 for $N_{max}=22$ with using the low-momentum
equivalent interaction of $\Lambda=3$ fm$^{-1}$. Results from the CD-Bonn
(AV18) potential are shown by thick (thin) curves.

The full CCSD results are seen to be almost independent on the oscillator
constant both for ground-state energies and radii of $^4$He, except for
the region below $\hbar\omega \simeq 20$ for $N_{max}=10$.
The tendency in the case of $N_{max}=10$ that the energy with
larger $\hbar\omega$ provides slightly lower energy
is understandable, because the s.p. wave functions of the larger
oscillator constant contain relatively larger high-momentum components.
This point, however,
needs to be clarified in detail together with the slight convex shape of the
oscillator constant dependence of the energy.
The difference of the solid and the long-dashed curves in Figs. 8$\sim$11
shows that the neglect of the potentials $u_{12}^{A,B}$ does not much affect
the energy and the radius. The ground-state energy becomes less attractive
by about 0.5 MeV and the radius becomes slightly smaller by about 0.02 fm
in the both cases of CD-Bonn and AV18.

The difference of the contributions of correlations described by  $u_{12}^{A,B}$
with CD-Bonn and AV18 reflects the different character of these interactions,
typically observed in the strength of the tensor components.
Together with the convergence with respect to the size of the model space,
we see that the many-body problem with the low-momentum equivalent
interaction is solved adequately by the CCM framework at the CCSD level.

If we neglect $s_1$ terms from the beggining, the CCD approximation,
calculated energies and radii vary with changing the oscillator constant
of the model space, as Figs. 8$\sim$11 show. Increasing the
frequency $\hbar \omega$, the energy attains a minimum value at
$\hbar\omega = 21\sim 23$ MeV and the radius decreases monotonically
in almost parallel with the r.m.s. radius of the reference state $|\Phi_0\rangle$,
Eq. (\ref{eq:hor}). Even in this case, the effect of $u_{12}^{A,B}$ is comparable
to the case of fully including $s_1$. It is interesting to observe that the
difference between the energy with $s_1=0$ at its minimum value and the full
CCSD energy is rather small. The r.m.s. radii with and without $s_1$ are also
seen to coincide each other at
the point where energies with and without $s_1$ become close.

The order of 0.5 MeV of the difference between calculated energies with and without
the $u_{12}^{A,B}$ terms is smaller than that of the energy gain due to the two-body
high-momentum correlations. This implies that the two-body correlations including
high-momentum components are primarily important and the remaining many-body
correlations summed up by $u_{12}^{A,B}$ in a restricted model space plays
a minor role in the magnitude of energy. This
is consistent with the conventional understanding, regarding the justification of the
independent particle picture in spite of the short-range repulsion of the bare
nucleon-nucleon interaction.

The above observation suggests that three-body correlations including
high-momentum components are worth to investigate quantitatively.
On the other hand, In the renormalization-group  view-point on which the
development of low-momentum interaction \cite{BOG} is conceptually based,
such correlations should be hidden in adjustable parameters as far as low-energy
properties of nuclei are concerned. When the experimental data at
intermediate energy in nuclear physics is accumulated in the
future, these correlations will be addressed as the important subject to
analyze.

\section{Summary}
We have evaluated the ground-state energy and radius of $^4$He in the
reformulated coupled-cluster method in which many-body average potentials
are introduced. This method was proposed by Suzuki in 1990's \cite{S92,SOK94},
but has not been practiced for actual nuclei. The formulation clarifies the
procedure to organize the Hamiltonian in a normal-ordered form with respect
to the reference state, and thus the structure of the CCM approach as
the many-body theory. The obtained decoupling equation in the
truncation at the two-body correlation level, commonly referred to as
the CCSD approximation, is arranged in a concise form as given
in Eq. (\ref{eq:dc12}) by virtue of the average potential, although the
content is necessarily not different from that of other CCM calculations
\cite{DH04}.   

Numerical calculations are carried out in the harmonic oscillator
basis. Major shells up to $N_{max}=22$ are prepared,
while the oscillator constant $\nu$ is varied in the range of $0.4\sim0.8$ fm$^{-2}$.
Besides the bare CD-Bonn and AV18 nucleon-nucleon potentials, we show
the results with using low-momentum equivalent interactions. The independence
of the results on the oscillator constant $\nu$ is demonstrated both for
the ground-state energy and the radius. When we ignore the $s_1$ transformation
amplitude, the energy shows the parabolic dependence on $\nu$ in the calculated
range and the radius decreases monotonically with increasing $\nu$. It is interesting
to see that the difference of the energies with and without $s_1$ is minimized
at around the point where the radii with and without $s_1$ coincide. 

It has been sometimes advocated \cite{BOG} that different bare interactions collapse to
universal low-momentum interaction. This property is well expected as far as
the diagonal matrix elements are concerned, because on-shell matrix
elements are directly related to experimental scattering data. However,
the difference of  inherent characters
of each bare interaction which are typically characterized by the strength
of tensor components persists in the low-momentum interaction.

There are several ways to extend the present stage of our calculations in the future.
The estimation of the contributions beyond the $s_1+s_{12}$ approximation
is necessary. The inclusion of $s_{123}$, however, affects the energy only through the
change of $s_1$ and $s_{12}$, when only two-body interactions are present.
Therefore, the net effect of those higher correlations may be small.
The estimation of the contribution of a three-nucleon interaction is
important to quantitatively describe the saturation properties of nuclei.
The expression for the contribution of the three-nucleon force to the ground-state
energy is presented in Appendix D, which helps to understand the different character
of the contributions of the three-nucleon force and three-body correlations
(or an induced three-nucleon force).
As for the treatment of short-range singularity of the nucleon-nucleon interaction,
it is desirable to renormalize the high-momentum component of the
bare nucleon-nucleon interaction in a more seamless way, working always
in a harmonic oscillator basis. Finally, numerical calculations should be
extended to larger nuclei and also the consideration of excited states.

\acknowledgments
This work is supported by Grant-in Aids for Scientific Research (C) from
the Japan Society for the Promotion of Science (Grant No. 22540288).
The authors are grateful to K. Suzuki and H. Kamada for useful comments.

\newpage
\begin{widetext}
\appendix
\section{Similarity transformation of the one-body operator}
Remembering that the product of  more than two transformation amplitudes
in which the same suffix appears, e.g. $s_1s_{12}$, vanishes,
the similarity-transformation of the one-body operator $\sum_i t_i$
leads to up to 4-body operators.
The expansion of the similarity-transformation $e^{-S}\sum_i t_i e^S$ with
$S=\sum_i s_i+\sum_{i<j}s_{ij}+ \sum_{i<j<k}s_{ijk}$ provides the
following terms:
\begin{eqnarray}
\mbox{1-body part} & & \sum_i (1-s_i)t_i(1+s_i),\\
\mbox{2-body part} & & \sum_{i<j}\{(t_i+t_j)s_{ij}-s_{ij}(t_i+t_j)
 -(s_it_i+s_jt_j) s_{ij}- s_{ij}(t_is_i+t_js_j)\},\\
\mbox{3-body part} & & -\sum_{i,j,k(j\ne k)} s_{\widehat{ij}}t_is_{\widehat{ik}}
-\sum_{i,(j<k)} \{ s_it_i s_{\widehat{ijk}}+ s_{\widehat{ijk}} t_is_i \},\\
\mbox{4-body part} & &
-\sum_{i,j,(k<\ell)}(s_{\widehat{ik\ell}}t_i s_{\widehat{ij}}+s_{\widehat{ij}}t_is_{\widehat{ik\ell}}),
\end{eqnarray}
where the widehat $\widehat{ijk}$ in the suffix
means that $i$, $j$ and $k$ are to be arranged in an ascending order.

\section{Similarity transformation of the two-body interaction}
The expansion of the similarity-transformation $e^{-S}\left(\sum_{i<j}v_{ij}\right)e^S$
provides up to 6-body operators. Explicit expressions are the following,
omitting terms including $s_{ijk\ell}$ and higher excitation operators.
Note that
the two-body part corresponds to $\sum_{i<j} (\tilde{v}_{ij}-\tilde{u}_{ij})$ in Eq. (\ref{eq:stf}),
the three-body part to $\sum_{i<j<k} (\tilde{v}_{ijk}-\tilde{u}_{ijk})$ originating from
the two-body interaction $v_{ij}$, and so on.
\begin{eqnarray}
\mbox{2-body part}  & & \sum_{i<j} (1-s_i-s_j+s_is_j-s_{ij})v_{ij}(1+s_i+s_j+s_is_j+s_{ij}),\\
\mbox{3-body part} & & \sum_{i<j} \sum_k \{(1-s_i-s_j+s_is_j-s_{ij})v_{ij}
(s_{\widehat{ik}}+s_{\widehat{jk}} +s_{\widehat{ik}} s_j+s_{\widehat{jk}}s_i+s_{\widehat{ijk}})
 \nonumber \\
 & & -(s_{\widehat{ik}}+s_{\widehat{jk}}-s_{\widehat{ik}}s_j
-s_{\widehat{jk}}s_i+s_{\widehat{ijk}})v_{ij} (1+s_i+s_j+s_is_j+s_{ij})\},\\
\mbox{4-body part}     & & \sum_{i<j} \sum_{k<\ell} \{(1-s_i-s_j-s_{ij}+s_is_j)v_{ij}
(s_{\widehat{ik\ell}}+s_{\widehat{jk\ell}}
 +s_is_{\widehat{jk\ell}}+s_js_{\widehat{ik\ell}}) \nonumber \\
 & &-(s_{\widehat{ik\ell}}+s_{\widehat{jk\ell}}
 -s_is_{\widehat{jk\ell}}-s_js_{\widehat{ik\ell}})v_{ij}(1+s_i+s_j+s_{ij}+s_is_j)\} \nonumber \\
 & &+\sum_{i<j}\sum_{k\ne\ell}\{(1-s_i-s_j-s_{ij}+s_is_j)v_{ij}s_{\widehat{ik}}s_{\widehat{j\ell}}
 +s_{\widehat{ik}}s_{\widehat{j\ell}}v_{ij}(1+s_i+s_j+s_{ij}+s_is_j) \nonumber \\
 & &-(s_{\widehat{ik}}+s_{\widehat{jk}}+s_{\widehat{ijk}}-s_is_{\widehat{jk}}
 -s_js_{\widehat{ik}})v_{ij}(s_{\widehat{i\ell}}+s_{\widehat{j\ell}}+s_{\widehat{ij\ell}}
 +s_is_{\widehat{j\ell}}+s_js_{\widehat{i\ell}})\},\\
 \mbox{5-body part} & & \sum_{i<j}\sum_{k< \ell}\sum_m \{(1-s_i-s_j-s_{ij}+s_is_j)v_{ij}
(s_{\widehat{ik\ell}}s_{\widehat{jm}}+s_{\widehat{jk\ell}}s_{\widehat{im}})  \nonumber \\
 & &+(s_{\widehat{ik\ell}}s_{\widehat{jm}}+s_{\widehat{jk\ell}}s_{\widehat{im}})
 v_{ij}(1+s_i+s_j+s_{ij}+s_is_j) \nonumber \\
 & &-(s_{\widehat{im}}+s_{\widehat{jm}}+s_{\widehat{ijm}}-s_is_{\widehat{jm}}-s_js_{\widehat{im}})
 v_{ij}(s_{\widehat{ik\ell}}+s_{\widehat{jk\ell}}+s_is_{\widehat{jk\ell}}+s_js_{\widehat{ik\ell}})
 \nonumber \\
 & &-(s_{\widehat{ik\ell}}+s_{\widehat{jk\ell}}-s_is_{\widehat{jk\ell}}-s_js_{\widehat{ik\ell}})v_{ij}
(s_{\widehat{im}}+s_{\widehat{jm}}+s_{\widehat{ijm}}+s_is_{\widehat{jm}}+-s_js_{\widehat{im}})\} 
  \nonumber \\
 & & +\sum_{i<j}\sum_{k\ne\ell\ne m}
 \{-(s_{\widehat{ik}}+s_{\widehat{jk}}+s_{\widehat{ijk}}-s_is_{\widehat{jk}}-s_js_{\widehat{ik}})
 v_{ij}s_{\widehat{i\ell}}s_{\widehat{jm}}\nonumber \\
 & & +s_{\widehat{i\ell}}s_{\widehat{jm}}v_{ij}
 (s_{\widehat{ik}}+s_{\widehat{jk}}+s_{\widehat{ijk}}-s_is_{\widehat{jk}}-s_js_{\widehat{ik}})\},\\
 \mbox{6-body part} & &  \sum_{i<j} \sum_{k<\ell}\sum_{m<n} \{(1-s_i-s_j-s_{ij}+s_is_j)v_{ij}
 s_{\widehat{ik\ell}}s_{\widehat{jmn}}+s_{\widehat{ik\ell}}s_{\widehat{jmn}}v_{ij}
 (1+s_i+s_j+s_{ij}+s_is_j)\nonumber\\
 & & -(s_{\widehat{ik\ell}}+s_{\widehat{jk\ell}}-s_is_{\widehat{jk\ell}}-s_js_{\widehat{ik\ell}})
 v_{ij}(s_{\widehat{imn}}+s_{\widehat{jmn}}+s_is_{\widehat{jmn}}+s_js_{\widehat{imn}})\}\nonumber\\
 & & +\sum_{i<j}\sum_{k,\ell}\sum_{m<n}
 \{(s_{\widehat{i\ell}}s_{\widehat{jmn}}+s_{\widehat{j\ell}}s_{\widehat{imn}})v_{ij}
 (s_{\widehat{ik}}+s_{\widehat{jk}}+s_{\widehat{ijk}}+s_is_{\widehat{jk}}
 +s_js_{\widehat{ik}})\nonumber\\
 & & -(s_{\widehat{ik}}+s_{\widehat{jk}}+s_{\widehat{ijk}}-s_is_{\widehat{jk}}-s_js_{\widehat{ik}})
 v_{ij}(s_{\widehat{i\ell}}s_{\widehat{jmn}}+s_{\widehat{j\ell}}s_{\widehat{imn}})\nonumber\\
 & & +s_{\widehat{ik}}s_{\widehat{j\ell}}v_{ij}
 (s_{\widehat{imn}}+s_{\widehat{jmn}}+s_is_{\widehat{jmn}}+s_js_{\widehat{imn}})
  -(s_{\widehat{imn}}+s_{\widehat{jmn}}-s_is_{\widehat{jmn}}-s_js_{\widehat{imn}})
  v_{ij}s_{\widehat{ik}}s_{\widehat{j\ell}}\}\nonumber\\
 & & + \sum_{i<j}\sum_{k\ne\ell\ne m \ne n}
 s_{\widehat{ik}}s_{\widehat{j\ell}}v_{ij}s_{\widehat{im}}s_{\widehat{jn}}.
\end{eqnarray}

\section{Similarity transformation of three-nucleon interaction}
A three-body interaction $v_{ijk}$ give the following three-body terms
in the similarity-transformation. More than 4-body operators
are not shown for the sake of simplicity.
\begin{eqnarray}
& &\sum_{i<j<k}(1-s_i-s_j-s_k-s_{ij}-s_{jk}-s_{ik}+s_is_j+s_is_k+s_js_k-s_{ijk}
      +s_is_{jk}+s_js_{ik}+s_ks_{ij}-s_is_js_k+s_{ijk}) \nonumber \\
 & \times &v_{ijk}(1+s_i+s_j+s_k+s_{ij}+s_{jk}+s_{ik}
      +s_is_j+s_is_k+s_js_k+s_{ijk}+s_is_{jk}+s_js_{ik}+s_ks_{ij}+s_is_js_k+s_{ijk}).
\end{eqnarray}

\section{Expression of total energy with three-nucleon interaction}
When the Hamiltonian contains a three-nucleon interaction, $\sum_{i<j<k}v_{ijk}$,
the energy obtained from Eq. (\ref{eq:ccme}) becomes
\begin{eqnarray}
 E_0 &=& \sum_h \langle h|t_1(1+s_1)|h\rangle
 +\frac{1}{2}\sum_{hh'} \left\{ \langle hh'|v_{12}|hh'\rangle_A
 +\sum_p 2\times \langle hh'|v_{12}|ph'\rangle_A \langle p|s_1|h\rangle \right.
  \nonumber\\
 & & +\frac{1}{2}\sum_{pp'} 2\times \langle hh'|v_{12}|pp'\rangle_A
 \langle p|s_1|h\rangle \langle p'|s_1|h'\rangle
+\frac{1}{2}\sum_{pp'} \langle hh'|v_{12}|pp'\rangle_A
 \langle pp'|s_{12}|hh'\rangle \} \nonumber \\
 & & +\frac{1}{6}\sum_{hh'h''} \{\langle hh'h''|v_{123}(1+s_{123})|hh'h''\rangle_A
 + 3 \sum_{p} \langle hh'h''|v_{123}|ph'h''\rangle_A \langle p|s_1|h\rangle
  \nonumber \\
 & & +\frac{1}{2}\sum_{pp'} 6\times \langle hh'h''|v_{123}|pp'h''\rangle_A
 \langle p|s_1|h\rangle \langle p'|s_1|h'\rangle
 +\frac{1}{2}\sum_{pp'}3\times\langle hh'h''|v_{123}|pp'h''\rangle_A
 \langle pp'|s_{12}|hh'\rangle_A \nonumber \\
 & & \left. + \frac{1}{6}\sum_{pp'p''} 9\times\langle hh'h''|v_{123}|pp'p''\rangle_A
 \langle pp'|s_{12}|hh'\rangle_A \langle p''|s_1|h''\rangle \right\}.
\end{eqnarray}
It is useful to introduce the two-body interaction $v_{12(3)}$ by holding the
one coordinate of $v_{123}$ with occupied states,
\begin{equation}
 \langle a_1a_2|v_{12(3)}| a_3a_4\rangle_A \equiv \sum_{h} \langle a_1a_2h|
  v_{123}|a_3a_4h\rangle_A.
\end{equation}
Using this notation, the energy $E_0$ is written in a concise form as
\begin{eqnarray}
 E_0 &=& \sum_h \langle h|t_1(1+s_1)|h\rangle
 +\frac{1}{2}\sum_{hh'} \left\{ \langle hh'|v_{12}+\frac{1}{3}v_{12(3)}|hh'\rangle_A
 +\sum_p 2\times \langle hh'|v_{12}
 +\frac{1}{2}v_{12(3)}|ph'\rangle_A \langle p|s_1|h\rangle \right. \nonumber \\
 & & \left. +\frac{1}{2}\sum_{pp'} 2\times \langle hh'|v_{12}+v_{12(3)}|pp'\rangle_A
 \langle p|s_1|h\rangle \langle p'|s_1|h'\rangle
+\frac{1}{2}\sum_{pp'} \langle hh'|v_{12}+v_{12(3)}|pp'\rangle_A
 \langle pp'|s_{12}|hh'\rangle_A \right\} \nonumber \\
 & & +\frac{1}{6}\sum_{hh'h''} \left\{\langle hh'h''|v_{123}s_{123}|hh'h''\rangle_A
 + \frac{1}{6}\sum_{pp'p''} 9\times\langle hh'h''|v_{123}|pp'p''\rangle_A
 \langle pp'|s_{12}|hh'\rangle_A \langle p''|s_1|h''\rangle \right\}.
\end{eqnarray}
The factors $\frac{1}{3}$ and $\frac{1}{2}$ in front of $v_{12(3)}$ in the second
and third terms come from the statistical weight and are naturally derived.
The necessity of these factors are recently pointed out by Hebeler and
Schwenk in Ref. \cite{HS10}. It is noted, however, we also have to
consistently include $v_{ijk}$ and $s_{ijk}$ in decoupling equations.
The expressions become complicated, which are beyond the scope of this article. 
\bigskip
\end{widetext}


\begin{thebibliography}{99}
\bibitem{ME11} R. Machleidt and D.R. Entem, Phys. Rep. {\bf 503}, 1 (2011).
\bibitem{E06} E. Epelbaum, Prog. Part. Nucl. Phys. {\bf 57}, 654 (2006).
\bibitem{BOG} S.K. Bogner, T.T.S. Kuo, and A. Schwenk, Phys. Rep. {\bf 386}, 1 (2003).
\bibitem{MC} S. Pieper and R. wiringa, Ann. Rev. Nucl. Part. Sci. {\bf 51}, 53 (2001).
\bibitem{NCSM} P. Navr\'{a}til, S. Quaglioni1, I. Stetcu, and B.R. Barrett, J. Phys. G:
Nucl. Part. Phys. {\bf 36}, 083101 (2009).
\bibitem{DH04} D.J. Dean and M. Hjorth-Jensen, Phys. Rev. {\bf C69}, 054320 (2004).
\bibitem{C58} F. Coester, Nucl. Phys. {\bf 7}, 421 (1958).
\bibitem{CK60} F. Coester and H. K\"{u}mmel, Nucl. Phys. {\bf 17}, 477 (1960).
\bibitem{KLZ78} H. K\"{u}mmel, K.H. L\"{u}hrmann, and J.G. Zabolitzky, Phys. Rep. {\bf 36},
1 (1978).
\bibitem{BM07} R.J. Bartlett and M. Musial, Rev. Mod. Phys. {\bf 79}, 291 (2007).
\bibitem{SB09} I. Shavitt and R.J. Bartlett, {\it "Many-Body Methods in Chemistry and Physics"},
(Cambridge University Press, 2009).
\bibitem{HM99} J.H. Heisenberg and B. Mihaila, Phys. Rev. {bf C59}, 1440 (1999).
\bibitem{KD04} K. Kowalski, D.J. Dean, M. Hjorth-Jensen, T. Papenbrock, and P. Piecuch, 
Phys. Rev. Lett. {\bf 92}, 132501 (2004).
\bibitem{HD07} G. Hagen, D.J. Dean, M. Hjorth-Jensen, T. Papenbrock, and A. Schwenk,
Phys. Rev. {\bf C76}, 044305 (2007).
\bibitem{HP09} G. Hagen, T. Papenbrock, D.J. Dean, M. Hjorth-Jensen, and B. VelamurAsokan,
Phys. Rev. {\bf C80}, 021306 (2009).
\bibitem{HP10} G. Hagen, T. Papenbrock, D.J. Dean, and M. Hjorth-Jensen,
Phys. Rev. {\bf C82}, 034330 (2010).
\bibitem{JH11} G.R. Jansen, M. Hjorth-Jensen, G. Hagen, and T. Papenbrock,
Phys. Rev. {\bf C83}, 054306 (2011).
\bibitem{HP07} G. Hagen, T. Papenbrock, D.J. Dean, A. Schwenk, A. Nogga, M. W{\l}och,
and P. Piecuch, Phys. Rev. {\bf C76}, 034302 (2007).
\bibitem{S92} K. Suzuki, Prog. Theor. Phys. {\bf 87}, 937 (1992).
\bibitem{CDB} R. Machleidt, Phys. Rev. {\bf C63}, 024001 (2001).
\bibitem{AV18} R.B. Wiringa, V.G.J.Stoks, and R. Schiavilla, Phys. Rev. {\bf C51}, 38 (1995).
\bibitem{CZ66} J. \v{C}i\v{z}ek, J. Chem. Phys. {\bf 45}, 4256 (1966).
\bibitem{SOK94} K. Suzuki, R. Okamoto, and H. Kumagai, Nucl. Phys. {\bf A580}, 213 (1994).
\bibitem{NOG02} A. Nogga, H. Kamada, W. Glockle, and B.R. Barrett, Phys. Rev. {\bf C65},
054003 (2002). 
\bibitem{BLM54} K.A. Brueckner, C.A. Levinson, and H.M. Mahmoud, Phys. Rev. {\bf 95}, 217 (1954).
\bibitem{BR67} B.H. Brandow, Rev. Mod. Phys. {\bf 39}, 771 (1967).
\bibitem{SO94} K. Suzuki and R. Okamoto, Prog. Theor. Phys. {\bf 92}, 1045 (1994).
\bibitem{S08} I. Sick, Phys. Rev. {\bf C77}, 041302(R) (2008).
\bibitem{HS10} K. Hebeler and A. Schwenk, Phys. Rev. {\bf C82}, 014314 (2010).
\end{thebibliography}
\end{document}